\documentclass[aps,twocolumn]{revtex4-2}

\usepackage{graphicx}
\usepackage{lipsum}

\usepackage{dcolumn}
\usepackage{bm}
\usepackage{multirow}
\usepackage{float}
\usepackage{amsmath}
\usepackage{wrapfig}
\usepackage[normalem]{ulem}
\usepackage{color}
\usepackage{xr}
\usepackage{booktabs}
\usepackage{longtable}
\usepackage[colorlinks=true, citecolor=blue, urlcolor=black]{hyperref}
\usepackage{relsize}
\usepackage{bbm}
\usepackage{dsfont}
\usepackage{xcolor}

\begin{document}

\title{Valley-polarized Orbital and Spin Magnetism Induced by Femtosecond Optical Pulses in Two-Dimensional Semiconductors}

\author{M. S. Mrudul}
\email[]{mrudul.muraleedharan@physics.uu.se}
\affiliation{%
Department of Physics and Astronomy, P.\,O.\ Box 516, Uppsala University, S-75120 Uppsala, Sweden}

\author{Peter M. Oppeneer}
\affiliation{%
Department of Physics and Astronomy, P.\,O.\ Box 516, Uppsala University, S-75120 Uppsala, Sweden}

\date{\today}
		
\begin{abstract}
	We theoretically investigate the ultrafast generation of spin and orbital magnetism in a two-dimensional gapped Dirac system with spin-orbit coupling. { This system is representative of} two-dimensional hexagonal semiconductors, 
    such as 
    transition-metal dichalcogenides
    { that} exhibit valley-selective optical selection rules arising from the valley-contrasting magnetic texture of their band structure. Using a time-dependent density-matrix formalism, we demonstrate that circularly polarized laser pulses generate nonequilibrium magnetization under both resonant and multiphoton resonant conditions. We show that the induced spin and orbital magnetic moments can be distinctly controlled via the photon energy and polarization of the driving field. Furthermore, spin and orbital dynamics originate from fundamentally different light-matter coupling mechanisms, leading to qualitatively dissimilar temporal behaviors. The orbital magnetic moment couples directly to the external electric field, resulting in faster dynamics and pronounced Rabi-like oscillations, whereas the spin response develops gradually through spin-orbit coupling. Consequently, orbital dynamics is significantly more sensitive to electron-hole dephasing than the spin response. Our results highlight the importance of properly accounting for orbital contributions in future technologies that utilize femtosecond control of magnetism.
\end{abstract}

\keywords{ultrafast magnetization, orbital magnetic moment, spin magnetic moment, circularly polarized laser, inverse Faraday effect, valley polarization, nonlinear optics, spin--orbit coupling}

\maketitle

\section{Introduction}
Conventional magnetic control through external magnetic fields is often limited by relatively slow switching speeds and high energy dissipation, which motivated the search for alternative control mechanisms. Following the discovery of all-optical helicity-dependent magnetization switching~\cite{kimel2005ultrafast}, optical manipulation of magnetism using femtosecond circularly polarized light has become a promising route for ultrafast magnetization control~\cite{hansteen2006nonthermal,stanciu2007all,vahaplar2009ultrafast,satoh2010spin,kirilyuk2010ultrafast,vahaplar2012all}. A key mechanism underlying these effects is the perturbative nonlinear optical effect known as the inverse Faraday effect ~\cite{van1965optically,pershan1966theoretical}. In this process, circularly polarized light generates an effective optomagnetic field that can induce a net magnetization even in nonmagnetic systems~\cite{hertel2006theory,popova2011theory,popova2012theoretical,battiato2014quantum,berritta2016ab,adamantopoulos2025light}. However, all these works were mainly focused on coherent light-induced magnetization on metallic systems. Recent advances in strong-field physics in solids have revealed the emergence of novel optical phenomena in semiconductors, dielectrics, and semimetals~\cite{goulielmakis2007attosecond,ghimire2014strong,kruchinin2018colloquium,ghimire2019high}. In this context, understanding how the non-perturbative light-matter interaction influences coherent laser-induced magnetization in nonmagnetic nonmetallic materials represents an important and largely unexplored direction~\cite{okyay2020resonant,neufeld2023attosecond,neufeld2025linearly}.

Orbital and spin degrees of freedom are the fundamental building blocks of magnetization in matter. In solids, orbital magnetization is often neglected because the crystal electric field tends to quench the orbital angular momentum~\cite{kittel1955solid}. However, this assumption breaks down under non-equilibrium conditions. In particular, calculations of the inverse Faraday effect~\cite{berritta2016ab,adamantopoulos2025light}, as well as recent studies of the orbital Hall effect~\cite{go2018intrinsic,go2021orbitronics,salemi2022first}, have shown that orbital contributions can be substantial. It is therefore crucial to understand coherent laser-induced mechanisms that generate both spin and orbital magnetization, and to identify their fundamental differences on femtosecond timescales.

Two-dimensional materials, such as transition-metal dichalcogenides (TMDs), have attracted significant attention due to their remarkable optoelectronic properties~\cite{mak2010atomically,mak2016photonics,manzeli20172d}. In these materials, circularly polarized light can selectively excite electrons at a specific valley at the Brillouin zone edge, thereby introducing a controllable degree of freedom beyond charge and spin. Recently, many strong-field studies have been focused on ultrafast control of valley-degree of freedom in two-dimensional materials~\cite{langer2018lightwave,jimenez2020lightwave,mrudul2021light,rana2023all,mitra2024light,tyulnev2024valleytronics,gucci2026encoding}. Importantly, in TMDs, the valley-selective optical selection rules are directly linked to the orbital moment of the Bloch electrons~\cite{yao2008valley}. Furthermore, the bands near the valley points are spin-orbit split, resulting in spin-valley coupling~\cite{xiao2012coupled}. These unique properties make TMDs an ideal platform for exploring ultrafast magnetization dynamics induced by valley-polarized optical excitations.

Previously, Okyay \textit{et al.}\ investigated light-induced spin magnetization in TMDs driven by the inverse Faraday effect~\cite{okyay2020resonant}. In the present work, we extend this study by examining the laser-induced spin and orbital magnetization dynamics in a two-dimensional semiconductor with spin-orbit coupling. We explore the nonperturbative nonlinear optical mechanisms responsible for magnetization generation by solving the time-dependent quantum Liouville density-matrix equations. Furthermore, we demonstrate that spin and orbital magnetization exhibit distinct dynamical behaviors under strong-field excitation.

\section{Theoretical Methodology}
	{ 	
	{We employ a minimal tight-binding model for a two-dimensional gapped graphene system with spin--orbit coupling (SOC), following Ref.~\cite{okyay2020resonant}. The model captures the essential valley-contrasting physics of two-dimensional semiconductors such as TMDs. The details of the tight-binding model are described in Appendix~\ref{appendix:hk}. The electronic band structure { with the size of} 
    the spin and orbital magnetic moments { projected on the bands} are shown in Fig.~\ref{fig:bands}(a) and Fig.~\ref{fig:bands}(b), respectively. The conduction-band minima at the inequivalent K and K${'}$ points are treated as distinct valleys.  The system has a band gap ($\Delta_{\rm g}$) of 3.0~eV, and $\Delta_{\rm g}^{\rm sf} = 4.0$~eV denotes the electronic spin-flip excitation gap.}

	\begin{figure}[t]
		\includegraphics[width=\linewidth]{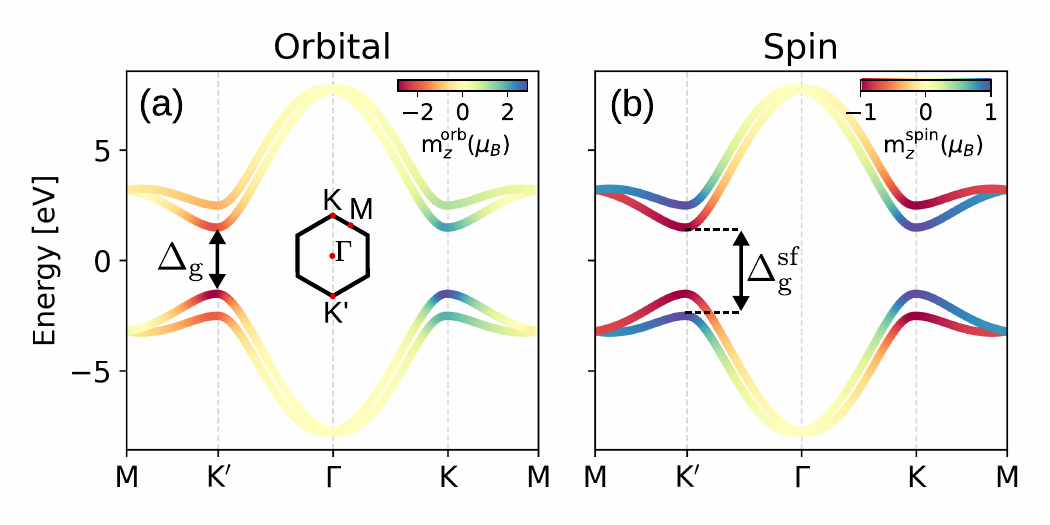}
		\caption{{Electronic band structure of an archetypal 2D semiconductor.}  {(a) Orbital and (b) spin components of magnetic moments ($m_{n\mathbf{k},z}$) projected onto the 
        band structure along the high-symmetry directions. The inset in (a) shows the high-symmetry points in the first Brillouin zone.}}
		\label{fig:bands}
	\end{figure}

	The spin magnetic moment of the Bloch electron in the $n^{th}$-band at \textbf{k} in the reciprocal space is defined as
	\begin{equation}
	\mathbf{m}^{\mathrm{spin}}_{n\mathbf{k}}
	=
	-g_s\frac{\mu_{\rm B}}{\hbar}
	\langle u_{n\mathbf{k}}|\hat{\mathbf{S}}|u_{n\mathbf{k}}\rangle ,
	\label{eq:spinmoment}
	\end{equation}
	where $|u_{n\textbf{k}\rangle}$ is the periodic part of the Bloch function, $\mu_{\rm B}$ is the Bohr magneton, $\hat{\mathbf{S}}$ is the spin operator, and $g_s \approx 2$ is the spin gyromagnetic factor.
	
	{In addition to the spin contribution, Bloch electrons possess an intrinsic orbital magnetic moment arising from the self-rotation of the electronic wave packet~\cite{chang1996berry,oppeneer1998magneto}. The orbital magnetic moment is given by 
	\begin{equation}
	\mathbf{m}^{\mathrm{orb}}_{n\mathbf{k}}
	=
	\frac{e}{2\hbar}
	\,\mathrm{Im}
	\left[
	\langle \partial_{\mathbf{k}} u_{n\mathbf{k}} |
	\times
	(\hat{\mathcal{H}}_{\mathbf{k}}-E_{n\mathbf{k}})
	|
	\partial_{\mathbf{k}} u_{n\mathbf{k}} \rangle
	\right],
	\label{eq:orbmoment}
	\end{equation}
	where $\hat{\mathcal{H}}_{\mathbf{k}}$ is the Bloch Hamiltonian, and $E_{n\mathbf{k}}$ is the eigen-energy of the Hamiltonian. }

	{The interaction of the material with the laser field is incorporated through the minimal coupling and within the dipole approximation as, $\mathcal{H}_{\mathbf{k}}(t) =
	\mathcal{H}\!\left(\mathbf{k}+e\mathbf{A}(t)\right)$,
	where $\mathbf{A}(t)$ is the vector potential related to the electric field through $\mathbf{E}(t)=-\frac{\partial }{\partial t}\mathbf{A}(t)$. The laser-excited electron dynamics { is} described using the single-particle density matrix formalism. The time evolution of the density matrix operator $\hat{\rho}_{\mathbf{k}}$ is governed by the quantum Liouville equation with decoherence~\cite{mrudul2024dependence},
	\begin{equation}
	\frac{d}{dt}\hat{\rho}_{\mathbf{k},mn}
	=
	-\frac{i}{\hbar}\left[
	\mathcal{H}_{\mathbf{k}}(t),
	\rho_{\mathbf{k}}
	\right]_{mn} - \frac{(1-\delta_{mn})}{T_2}\rho_{\mathbf{k},mn},
	\label{eq:densitymatrix}
	\end{equation}
	where $T_2$ is the electron-hole dephasing time. 
	Equation~(\ref{eq:densitymatrix}) is solved by projecting all operators onto the eigenbasis of the equilibrium Hamiltonian~\cite{mrudul2024dependence}. The time propagation is performed numerically using a fourth-order Runge--Kutta scheme with a time step of $0.02~\mathrm{fs}$. The Brillouin zone is sampled using a $200\times200$ $\mathbf{k}$-point mesh. We model laser pulses with a duration of $100\,\mathrm{fs}$ and a peak intensity of $10^{11}\,\mathrm{W/cm^2}$. Unless otherwise specified, all simulations are performed without including decoherence.}

	{We consider an ultrashort circularly polarized laser pulse with electric field
	\begin{equation}
	\mathbf{E}(t)
	=
	\frac{E_0 f(t)}{\sqrt{2}}
	\left[
	\hat{\mathbf{x}}\cos(\omega_0 t)
	\mp
	\hat{\mathbf{y}}\sin(\omega_0 t)
	\right],		
	\label{eq:laserfield}
	\end{equation}
	where \(E_0\) is the peak electric field amplitude, \(\omega_0\) is the carrier frequency, and \(f(t)\) is a \(\sin^2 (2\pi t/T)\) { ($0 \le t \le T/2$)} envelope function. The $-$ (+) sign corresponds to right (left) circular polarization. The peak laser intensity \(I_0\) and $E_0$ are related by $I_0 = \frac{1}{2} c \epsilon_0 E_0^2$.
	
	{The time evolution of the magnetization (net magnetic moment per unit cell) is calculated as
		\begin{equation}
		M_z(t)
		=
		\frac{1}{N_k}
		\sum_{\mathbf{k},mn}
		m_{\mathbf{k}nm,z}\,
		\rho_{\mathbf{k}mn}(t),
		\end{equation}
		where \(m_{\mathbf{k}nm,z}\) represents the matrix elements of either the spin or orbital magnetic moment operator along the \(z\)-direction, \(\rho_{\mathbf{k}mn}(t)\) is the time-dependent density matrix, and \(N_k\) denotes the total number of sampled \(\mathbf{k}\)-points in the Brillouin zone. The matrix elements of orbital magnetic moment { are} defined as ~\cite{bhowal2021orbital,pezo2022orbital,pozo2023multipole}
	\begin{equation}
	\begin{split}
	{\mathbf{m}}^{\rm orb}_{nn'}
	=&
	\frac{e}{2\hbar}
	\,\mathrm{Im}
	\left[
	\langle \partial_{\mathbf{k}} u_{n\mathbf{k}} |
	\times H_\mathbf{k} |
	\partial_{\mathbf{k}} u_{n'\mathbf{k}} \rangle
	\right] \\
	&-
	\frac{e}{4\hbar}
	\left[
	\epsilon_{n\textbf{k}}
	+
	\epsilon_{n'\mathbf{k}}
	\right]
	\mathrm{Im}
	\left[
	\langle \partial_{\mathbf{k}} u_{n\mathbf{k}} |
	\times
	|
	\partial_{\mathbf{k}} u_{n'\mathbf{k}} \rangle
	\right].
	\end{split}
	\end{equation}
    It is important to note that we regard the integrated magnetic moment per unit cell as the magnetization. This is an appropriate interpretation for spin magnetization, but it is not strictly valid for orbital magnetization. It has been shown that the latter contains an additional surface itinerant contribution~\cite{thonhauser2005orbital,xiao2005berry,shi2007quantum}. Nevertheless, the orbital magnetic moment remains physically meaningful, as it is directly related to the self-rotation of the wave packet~\cite{chang1996berry} and is also experimentally accessible through magnetic circular dichroism measurements~\cite{oppeneer1998magneto,Kunes2000,souza2008dichroic}.
}

\begin{figure}[t!]
	\centering
	\includegraphics[width=\linewidth]{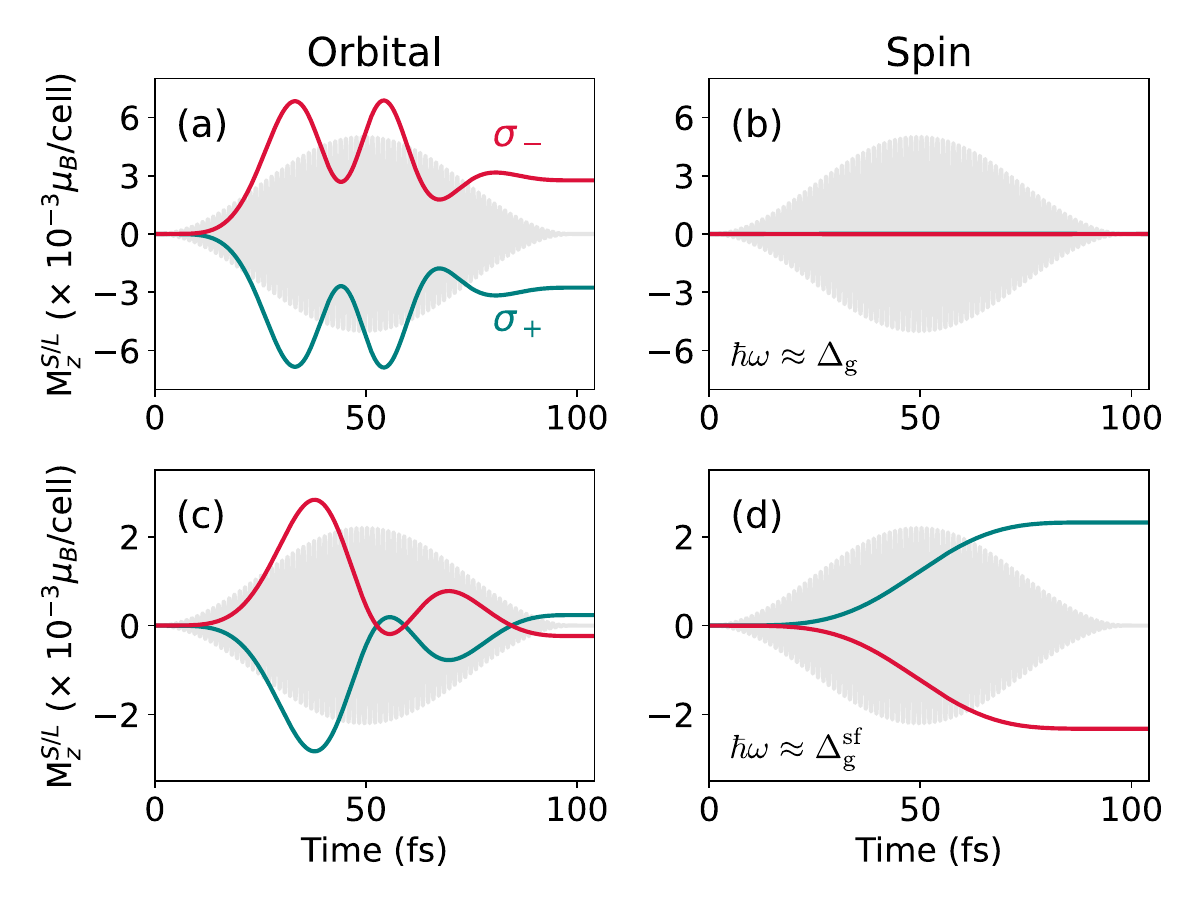}
	\caption{  Laser-driven dynamics of the orbital (left panels) and spin (right panels) magnetization. Red and blue curves correspond to magnetization induced by left ($\sigma_{-}$) and right ($\sigma_{+}$) circularly polarized laser pulses, respectively. The top panels show the response to a laser pulse with photon energy 3.1~eV ($\hbar\omega \approx \Delta_{\rm g}$), while the bottom panels correspond to excitation at 4.1~eV ($\hbar\omega \approx \Delta_{\rm g}^{\mathrm{sf}}$). The grey shaded regions indicate the temporal profile of the 100~fs long laser pulse with a peak intensity of $10^{11}$~W/cm$^{2}$.}
	\label{fig:figrt}
\end{figure}

\section{Results and Discussion}

The Hamiltonian preserves time-reversal symmetry but lacks inversion symmetry. Consequently, the magnetic moments satisfy the symmetry relation \(\mathbf{m}_{n}(-\mathbf{k}) = -\mathbf{m}_{n}(\mathbf{k})\), as evident from Figs.~\ref{fig:bands}(a) and \ref{fig:bands}(b). Nevertheless, the orbital and spin magnetic moments exhibit distinct textures in \(\mathbf{k}\)-space. As a result, optical excitation at different photon energies can induce markedly different orbital and spin magnetic responses. 

\subsection{Resonant excitation}

We demonstrate the dynamics of magnetization under resonant excitation in Fig.~\ref{fig:figrt}. Two excitation regimes are considered. In the first case, the photon energy is 3.1~eV, which lies above the band gap \(\Delta_{\rm g}\); the corresponding orbital and spin magnetization dynamics are shown in Figs.~\ref{fig:figrt}(a) and \ref{fig:figrt}(b), respectively. In the second case, the photon energy is 4.1~eV, which exceeds the spin-flip excitation gap \(\Delta_{\rm g}^{\mathrm{sf}}\); the corresponding orbital and spin magnetization dynamics are presented in Figs.~\ref{fig:figrt}(c) and \ref{fig:figrt}(d), respectively. We observe that the orbital contribution is comparable to, and in some cases even stronger than, the spin contribution. In all cases, the induced magnetization reverses sign when the helicity of the laser pulse is reversed. A key observation is that the remnant magnetization at the end of the laser pulse strongly depends on the photon energy. The remnant orbital magnetization dominates for excitation energies near \(\Delta_{\rm g}\), whereas the remnant spin magnetization dominates for excitation energies near \(\Delta_{\rm g}^{\mathrm{sf}}\) { ($\sim 4$ eV)}, { i.e., spin or orbital magnetization can selectively be induced.}

Interestingly, there is an apparent difference in the dynamics of orbital and spin magnetization in Fig.~\ref{fig:figrt}. The spin magnetization increases gradually and saturates after the peak of the laser pulse (Fig.~\ref{fig:figrt}(d)), whereas the orbital magnetization exhibits oscillations during the interaction with the laser pulse, with an oscillation frequency that is independent of the laser frequency (Figs.~\ref{fig:figrt}(a) and \ref{fig:figrt}(c)). As a result, the peak orbital magnetization during the dynamics differs from the remnant magnetization.  In addition, orbital magnetization increases on a faster time scale than its spin counterpart. In the following part, we show that this behavior is directly related to how light couples differently to orbital and spin degrees of freedom.

\begin{figure}[t]
	\centering
	\includegraphics[width=\linewidth]{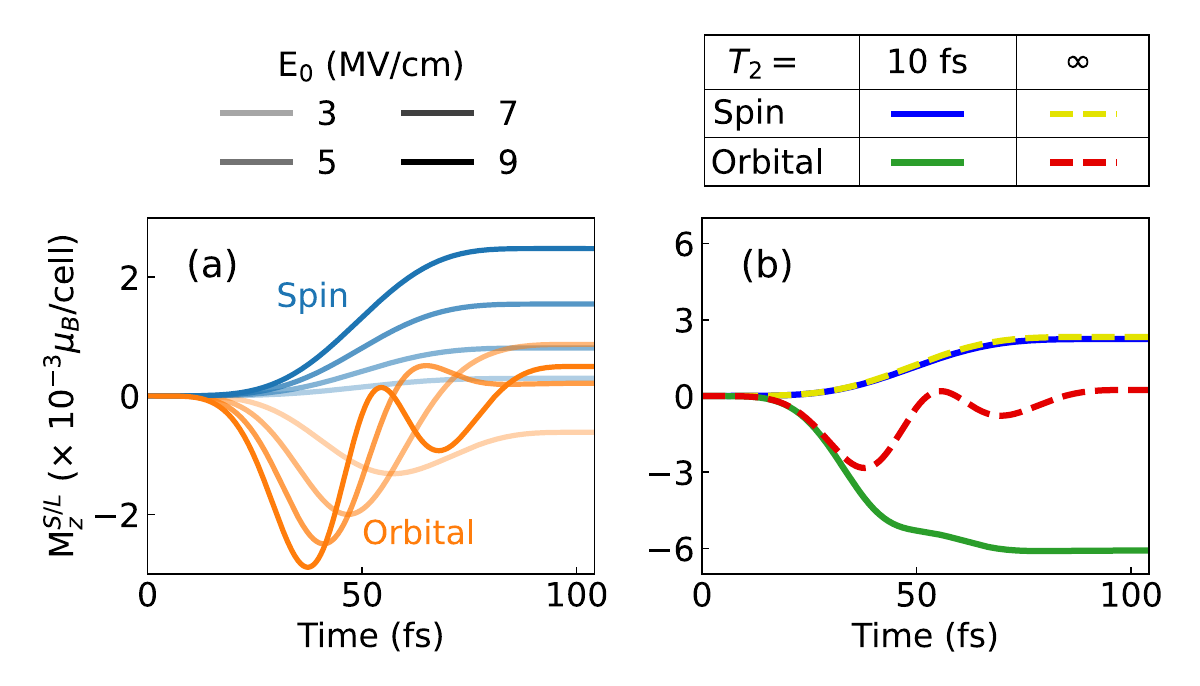}
	\caption{  (a) Laser-driven femtosecond dynamics of orbital (orange) and spin (blue) magnetization under laser pulses with { $\hbar \omega \approx \Delta_{\rm g}^{\rm sf}$ and} different electric field strengths, indicated by the transparency of the line plots. (b) Laser-driven dynamics of magnetization with an electron-hole dephasing time $T_2$ of 10~fs (blue for spin and green for orbital contributions) and in the absence of dephasing ($\infty$; yellow dashed for spin and red dashed for orbital). The peak laser intensity is $I_0 = 10^{11}$ W/cm$^2$, corresponding to the field amplitude $E_0 = 8.7$ MV/cm. }
	\label{fig:fig3}
\end{figure}

\begin{figure*}[t!]
	\centering
	\includegraphics[width=0.85\linewidth]{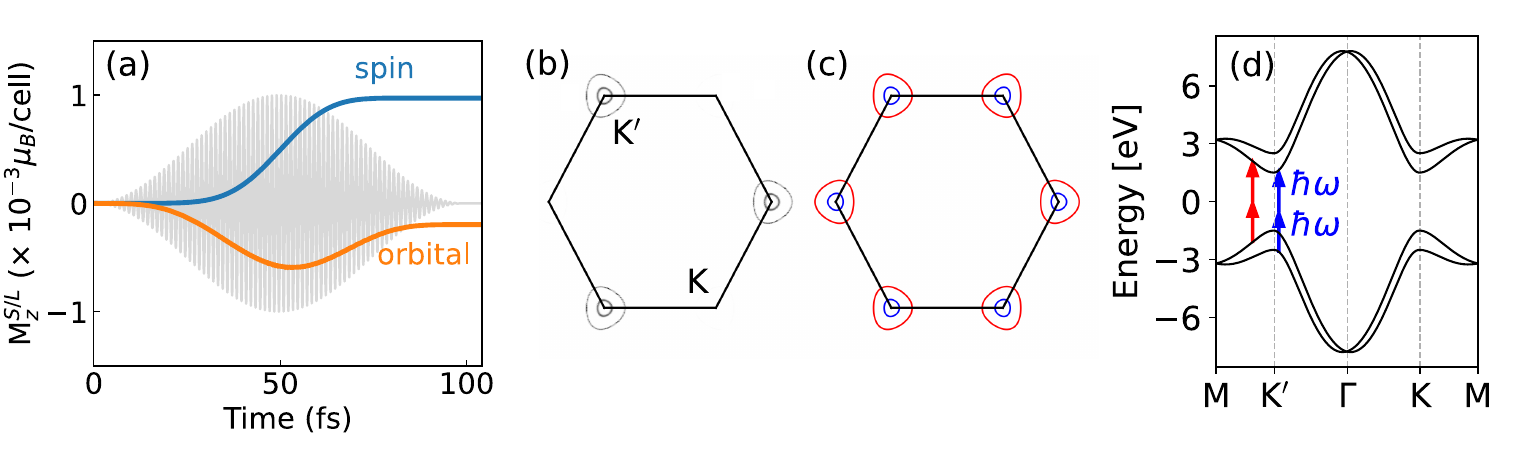}
	\caption{  (a) Laser-driven dynamics of the orbital (orange) and spin (light blue) magnetization for a laser pulse with photon energy $\hbar\omega = 2.1$~eV. (b) Electron population in the first conduction band (CB1) in reciprocal space at the end of the laser pulse. (c) Contours in the first Brillouin zone satisfying $E_{\mathrm{CB1},\mathbf{k}} - E_{\mathrm{VB1},\mathbf{k}} = 2\hbar\omega$ (red) and $E_{\mathrm{CB1},\mathbf{k}} - E_{\mathrm{VB2},\mathbf{k}} = 2\hbar\omega$ (blue). (d) Schematic illustration of two-photon excitation induced by a right-circularly polarized laser field. The population generated in the first conduction band arises from transitions from the first valence band (VB1\,$\rightarrow$\,CB1, blue arrows) and the second valence band (VB2\,$\rightarrow$\,CB1, red arrows).}
	\label{fig:lastfig}
\end{figure*}

The time evolution of the magnetization is governed by the equation of motion, $\frac{\partial}{\partial t}\langle m_z \rangle = -\frac{i}{\hbar}\langle[\hat{m}_z,\hat{\mathcal{H}}(t)]\rangle$. The total Hamiltonian can be decomposed as $\hat{\mathcal{H}}(t)
=
\hat{\mathcal{H}}_0
+
\hat{\mathcal{H}}_{\rm SOC}
+
\hat{\mathcal{H}}_{\rm lm}(t)$, 
where $\hat{\mathcal{H}}_0$ { includes the} single-electron kinetic energy and periodic potential, $\hat{\mathcal{H}}_{\rm SOC}$ represents the spin-orbit coupling, and $\hat{\mathcal{H}}_{\rm lm}(t) = -e \hat{\mathbf{r}} \cdot \mathbf{E}(t)$ is the light-matter interaction term.  In the case of spin dynamics, only the spin-orbit coupling contributes to the time evolution, implying that the spin couples to the external electric field only indirectly through $\hat{\mathcal{H}}_{\rm SOC}$. 
In contrast, for the orbital magnetization, all terms in the Hamiltonian contribute. This means that the characteristic timescale of spin dynamics is governed by $\hat{\mathcal{H}}_{\rm SOC}$, and therefore the spin degree of freedom exhibits slower dynamics compared to the orbital degree of freedom. As a consequence of the direct coupling between the orbital degrees of freedom and the external electric field, the orbital dynamics exhibits Rabi-like oscillations. Rabi oscillation is the coherent oscillation of population between two quantum states induced by a resonant external driving field, where the frequency of the oscillation is proportional to the coupling term~\cite{gerry2023introductory}. We analyze the implications of these effects in detail below.

{ Next,} we examine 
the regime where the laser frequency lies above $\Delta_{\rm g}^{\rm sf}$ [as shown in Figs.~\ref{fig:figrt}(c)-(d)], for which both spin and orbital dynamics become appreciable. In the following, we consider only left-circularly polarized light. Since the frequency of the Rabi-like oscillations is determined by the coupling strength of the external driving field, it can be effectively tuned by varying the electric field amplitude. To investigate this behavior, we present the field-strength dependence of the laser-induced magnetic dynamics in Fig.~\ref{fig:fig3}(a). The induced spin magnetization increases monotonically with increasing electric field strength, reflecting its indirect coupling to the driving field through spin-orbit interaction.  In contrast, the orbital dynamics responds more sensitively: changes in the electric field alter the oscillation frequency, leading to non-monotonic, more complex behavior.

So far, our analysis has been restricted to a closed quantum system, where electron-hole dephasing is neglected. In realistic systems, however, decoherence (see Eq.~(\ref{eq:densitymatrix})) is unavoidable because the electronic subsystem interacts with its environment, such as phonons or other electrons. Since Rabi oscillations rely on quantum coherence between the involved states, dephasing naturally suppresses these oscillations. Figure~\ref{fig:fig3}(b) compares the laser-driven spin and orbital dynamics for a { realistic} dephasing time of 10 fs with { those of} the decoherence-free case. A striking difference emerges: the spin dynamics remains nearly unchanged, whereas the orbital response is strongly altered. Dephasing suppresses the oscillatory character of the orbital motion and drives the system more rapidly toward saturation. As a result, the induced orbital magnetization becomes more robust { than}
the spin contribution, reaching a value nearly twice as large as the induced spin magnetization. This shows that the { resulting} 
orbital magnetization is governed not only by the driving field but also by the underlying decoherence processes during the electron dynamics. These differences highlight the fundamentally different roles of light-matter coupling in spin and orbital degrees of freedom under strong optical driving.

\subsection{Below band-gap excitation}

Up to this point, our discussion has been restricted to above-band-gap excitations. The density-matrix propagation scheme naturally incorporates processes of all orders in the electric field, whether resonant or non-resonant. We now investigate nonlinear optical processes induced by below-band-gap excitations that generate magnetization dynamics. Figure~\ref{fig:lastfig}(a) shows the dynamics of the spin (blue) and orbital (orange) magnetization driven by a left-circularly polarized laser pulse with photon energy 2.1~eV. Interestingly, substantial spin and orbital magnetization are generated, with magnitudes similar to those obtained under resonant excitation. Moreover, the spin and orbital contributions exhibit distinct dynamical behaviors, analogous to those observed in the resonant case in Fig.~\ref{fig:figrt}.

To further elucidate the origin of the induced magnetization, we analyze the electron population in the first conduction band (CB1) after the laser pulse, as shown in Fig.~\ref{fig:lastfig}(b). We observe a pronounced valley-polarized excitation, with carriers predominantly localized around the K$'$ valley. Interestingly, the population forms two concentric triangular rings in momentum space. We attribute this structure to { distinct} two-photon absorption pathways from different valence bands into CB1. For two-photon absorption, the relevant resonance condition is that the energy difference between the initial and final states satisfies $\Delta E = 2\hbar\omega$. To identify the contributing $\mathbf{k}$-points, we plot in Fig.~\ref{fig:lastfig}(c) the corresponding iso-energy contours satisfying this condition. The contours in Fig.~\ref{fig:lastfig}(c) show excellent agreement with the calculated population distribution in CB1 (Fig.~\ref{fig:lastfig}(b)). The double-ring structure in Fig.~\ref{fig:lastfig}(c) results from two distinct electronic transitions, as schematically illustrated in Fig.~\ref{fig:lastfig}(d). The outer ring (red) originates from two-photon transitions between the second valence band (VB2) and CB1, while the inner ring (blue) arises from transitions between the first valence band (VB1) and CB1. Notably, the VB1\,$\rightarrow$\,CB1 transition predominantly contributes to spin dynamics due to its spin-flip character (Fig.~\ref{fig:bands}(b)), whereas the VB2\,$\rightarrow$\,CB1 transition primarily governs the orbital response (Fig.~\ref{fig:bands}(a)). These results demonstrate that distinct two-photon excitation channels can selectively drive spin and orbital dynamics in different regions of momentum space. 

\begin{figure}[t!]
	\centering
	\includegraphics[width=\linewidth]{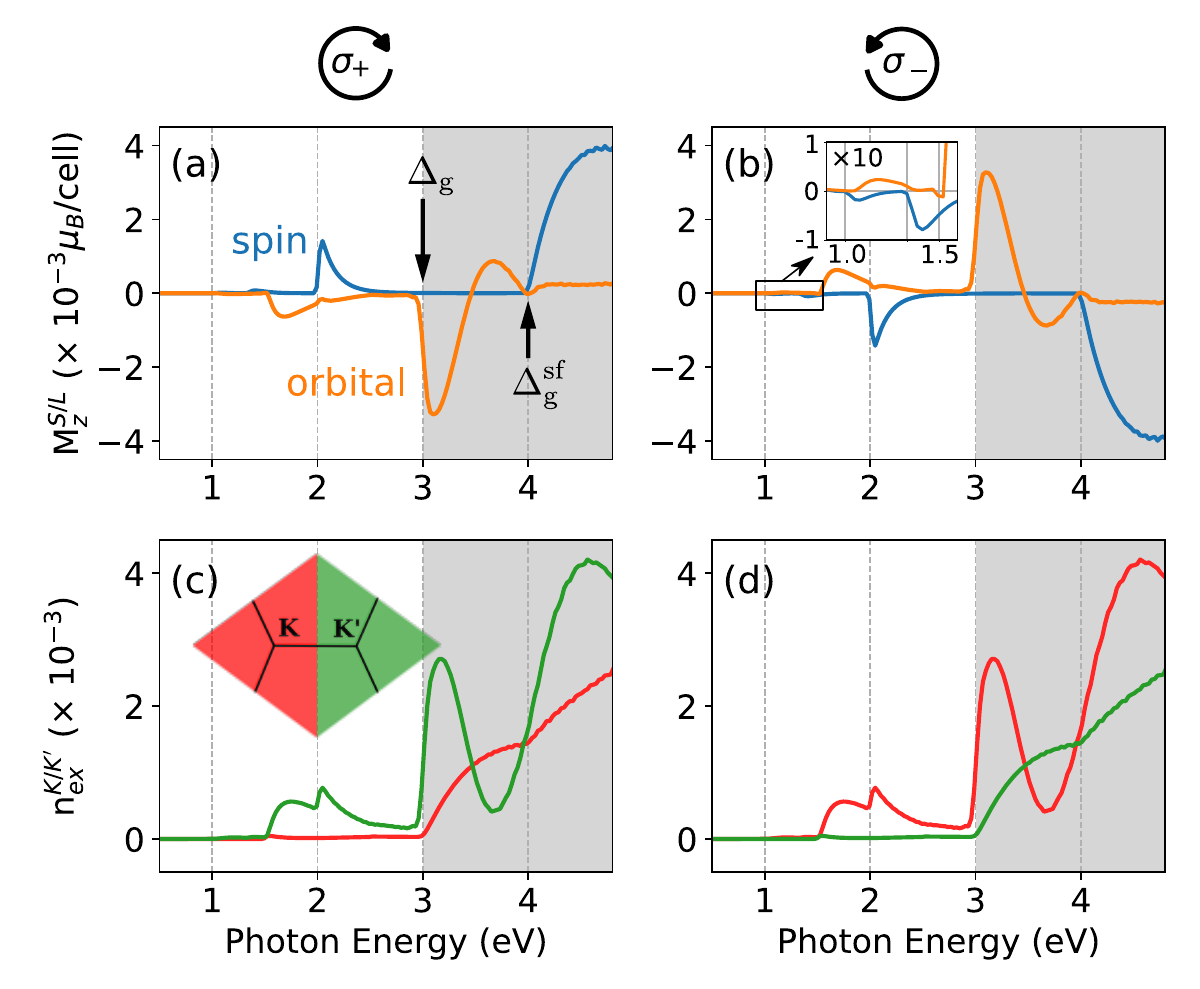}
	\caption{  Laser-induced spin (blue) and orbital (orange)  remnant magnetization  (calculated at the end of the laser pulse) as a function of the driving photon energy for (a) left- and (b) right-circularly polarized laser pulses. Panels (c) and (d) show the laser-excited electron population in the conduction bands at the end of the laser pulse, integrated around the K (red) and K$'$ (green) valleys, for (c) left- and (d) right-circularly polarized laser pulses. The separation of the reciprocal space unit-cell around the K and K$'$ valleys is illustrated in the inset of panel (c). The gray-shaded region indicates photon energies above the band gap, $\Delta_{\rm g}$.}
	\label{fig:figendepfinal}
\end{figure}

Having established that nonlinear below-band-gap excitations can generate significant spin and orbital dynamics, we now investigate this mechanism in greater detail and compare it with the response under above-band-gap excitation conditions. Figures~\ref{fig:figendepfinal}(a) and (b) show the laser-induced remnant magnetization as a function of the laser photon energy for right- and left-circularly polarized pulses, respectively. The grey-shaded region in Fig.~\ref{fig:figendepfinal} denotes the above-band-gap excitation regime. In this region, a pronounced peak in the orbital magnetic response appears for photon energies above $\Delta_{\rm g}$, while a strong spin magnetic response peak emerges above the spin-flip gap $\Delta_{\rm g}^{\mathrm{sf}}$. In addition to these single-photon excitation channels, we observe pronounced nonlinear responses below the band gap. In particular, the peak orbital response around $\hbar\omega \approx 1.5~\mathrm{eV} = \Delta_{\rm g}/2$ and the peak spin response around $\hbar\omega \approx 2~\mathrm{eV} = \Delta_{\rm g}^{\mathrm{sf}}/2$ originate from resonant two-photon absorption processes satisfying the condition $\Delta E = 2\hbar\omega$. Furthermore, the inset of Fig.~\ref{fig:figendepfinal}(b) reveals higher-order nonlinear processes, including a three-photon process inducing orbital magnetization at $\hbar\omega \approx 1~\mathrm{eV}$, as well as three-photon ($\hbar\omega \approx 1.33~\mathrm{eV}$) and four-photon ($\hbar\omega \approx 1~\mathrm{eV}$) processes contributing to the spin magnetic response. Interestingly, around $\hbar\omega \approx 1~\mathrm{eV}$, the four-photon-induced spin response and the three-photon-induced orbital response coexist with comparable magnitudes.

The spin-response peak at 2 eV is consistent with the spin magnetization reported in Ref.~\cite{okyay2020resonant}, where it was attributed to the inverse Faraday effect. 
The inverse Faraday effect is conventionally treated as a second-order nonlinear optical process, however, in metals the induced magnetization is typically associated with a difference-frequency term that does not lead to population transfer between states~\cite{berritta2016ab,kirilyuk2010ultrafast,battiato2014quantum}. In contrast, here the induced magnetization arises from a sum-frequency term (two-photon resonant excitation) that does involve population transfer~\cite{mrudul2024ab}. Nevertheless, it is important to note that the obtained laser-induced magnetization values are comparable to the reported values for metals~\cite{berritta2016ab,adamantopoulos2025light}.

Finally, we investigate the connection between the remnant magnetism and the valley-polarized conduction band population after the laser pulse, as shown in Figs.~\ref{fig:figendepfinal}(c) and \ref{fig:figendepfinal}(d). The conduction band population is integrated over regions surrounding the K and K$'$ valleys, as illustrated in the inset of Fig.~\ref{fig:figendepfinal}(c). Notably, both the induced spin and orbital magnetization are closely linked to valley-polarized electronic excitations in both resonant and off-resonant regimes. Moreover, reversing the laser helicity switches the valley polarization between K and K$'$, thereby inducing magnetization with opposite signs.

\section{Conclusion}
In conclusion, we investigated ultrafast laser-induced dynamics of magnetization in a two-dimensional gapped 
TMD model with spin-orbit coupling. Using a real-time density-matrix formalism that captures non-perturbative interactions, we demonstrated that circularly polarized laser pulses can generate substantial non-equilibrium magnetization through valley-selective electronic excitation. Spin { or} orbital magnetization can be selectively induced by tuning the photon energy. This energy selectivity reflects the underlying magnetic texture in momentum space.

Our results reveal fundamentally different temporal behaviors of { optically induced} spin and orbital magnetization. Spin dynamics evolves relatively slowly through indirect coupling { to the light field} mediated by spin-orbit interaction, whereas the orbital part couples directly to the external electric field and exhibits pronounced nonlinear effects, including Rabi-like oscillations. We further demonstrated that decoherence can strongly suppress these orbital oscillations, thereby \textit{enhancing} the { resulting} orbital response relative to its spin counterpart. 

{ Considering below band-gap excitation,} we demonstrated that strong spin and orbital magnetization can also be generated via multiphoton excitation. Finally, we established a direct correspondence between the { resultant} magnetization and valley-polarized conduction-band populations in both resonant and off-resonant excitation regimes. Overall, our findings highlight the crucial role of nonlinear light-matter interactions in controlling spin and orbital degrees of freedom on femtosecond timescales.

\begin{acknowledgments}
{ This work has been supported by the Swedish Research Council (VR), the German Research Foundation (Deutsche Forschungsgemeinschaft) through CRC/TRR 227 ``Ultrafast Spin Dynamics" (project MF, project-ID: 328545488), and the K.\ and A.\ Wallenberg Foundation (Grants No.\ 2022.0079 and 2023.0336).}
We { further} acknowledge funding from the European Union’s HORIZON EUROPE, under Grant Agreement No. 101129641, “OBELIX”. The calculations were partially supported by resources provided by the National Academic Infrastructure for Supercomputing in Sweden (NAISS) at NSC Linköping, partially funded by VR through Grant No. 2022-06725.
\end{acknowledgments}

\appendix

\section{Tight-binding Hamiltonian}
\label{appendix:hk}
{  We consider a generic two-dimensional hexagonal insulating system described by an effective tight-binding Hamiltonian, incorporating both sublattice inversion-symmetry breaking and spin-orbit coupling. The corresponding Bloch Hamiltonian is adopted from Ref.~\cite{okyay2020resonant}, and is written as 
\begin{equation}
\begin{split}
\hat{\mathcal{H}}(\mathbf{k})
=&
~ \hat{\mathcal{H}}_0(\mathbf{k})
+
\hat{\mathcal{H}}_{\mathrm{SOC}}(\mathbf{k}) \\
=&
~ \mathbf{h}(\mathbf{k}) \cdot \hat{\boldsymbol{\sigma}}
+
\hat{\sigma}_z \, \boldsymbol{\lambda}(\mathbf{k}) \cdot \hat{\mathbf{S}},
\label{eq:Hk}
\end{split}
\end{equation}
where $\hat{\boldsymbol{\sigma}}=(\hat{\sigma}_x,\hat{\sigma}_y,\hat{\sigma}_z)$ are Pauli matrices acting in the sublattice (pseudospin) space, and $\hat{\mathbf{S}}=(\hat{S}_x,\hat{S}_y,\hat{S}_z)$ are spin operators. The Hamiltonian acts in the direct product space of sublattice and spin degrees of freedom.

The first contribution in Eq.~(\ref{eq:Hk}) corresponds to the massive graphene Dirac Hamiltonian with
\begin{equation}
\mathbf{h}(\mathbf{k})
=
[h_x(\mathbf{k}),h_y(\mathbf{k}),\Delta].
\end{equation}
Here, $h_x(\mathbf{k})-ih_y(\mathbf{k})
=
-t_0\sum_{i=1}^{3}e^{i\mathbf{k}\cdot\boldsymbol{\tau}_i}$  with $t_0$ denoting the nearest-neighbor hopping amplitude and $\boldsymbol{\tau}_i$ the nearest-neighbor lattice vectors, and $\Delta$ is a sublattice-staggered potential induced by broken inversion symmetry.

The second term in Eq.~(\ref{eq:Hk}) describes the generalized SOC contribution written as
\begin{equation}
\boldsymbol{\lambda}(\mathbf{k})
=
[g_y(\mathbf{k}),-g_x(\mathbf{k}),\mu_z(\mathbf{k})].
\end{equation}
The terms proportional to $g_x(\mathbf{k})$ and $g_y(\mathbf{k})$ correspond to the intrinsic Rashba or pseudospin inversion asymmetry (PIA) SOC interaction, which gives rise to nontrivial spin dynamics because the spin operators no longer commute with the Hamiltonian.

The momentum-dependent SOC functions are defined as
\begin{equation}
\mu_z(\mathbf{k})
=
\frac{4}{3a^2}\lambda_{\rm I}
\sum_{i=1}^{3}
\sin(\mathbf{k}\cdot\boldsymbol{\nu}_i)d_{iz},
\end{equation}
and
\begin{equation}
g_j(\mathbf{k})
=
\frac{4}{3a}\lambda_{\mathrm{PIA}}
\sum_{i=1}^{3}
\sin(\mathbf{k}\cdot\boldsymbol{\nu}_i)\nu_{ij},
\end{equation}
where $a$ is the lattice constant, $\lambda_{\rm I}$ is the intrinsic SOC strength, and $\lambda_{\mathrm{PIA}}$ parametrizes the PIA SOC interaction. The vectors $\boldsymbol{\nu}_i$ correspond to next-nearest-neighbor hopping energies.

This effective model captures the essential low-energy physics of inversion-asymmetric two-dimensional materials such as TMDs. Throughout this work, we use the following parameters $t_0$ = 2.50$~\mathrm{eV}$, $\Delta$ = 2.00$~\mathrm{eV}$, $\lambda_{\rm I}$ = 0.50$~\mathrm{eV}$, $\lambda_{\mathrm{PIA}}$ = 0.50$~\mathrm{eV}$,
with lattice constant $a$ = 2.46$~\mathrm{\AA}$.


%

\end{document}